\documentclass[showkeys,showpacs,prd]{revtex4}
\usepackage{amsmath}
\usepackage{amssymb}
\usepackage{graphicx}

\begin{document}

\title{
Decomposed Finsler geometry: decomposed Finsler gravity and physical applications
}

\author{
Ascar K. Aringazin,$^{1}$
\footnote{Email: aringazin@gmail.com}
and
Vladimir Dzhunushaliev$^{1,2}$
\footnote{Email: v.dzhunushaliev@gmail.com}
}
\affiliation{$^1$
Institute for Basic Research, Eurasian National University, Astana, 010008, Kazakhstan, \\
$^2$
Dept. Theor. and Nucl. Phys., Kazakh National University, Almaty, 050038, Kazakhstan
}

\begin{abstract}
Special class of Finsler metrics that can be decomposed to the product of two Riemannian metrics is considered. Based on such decomposition a new kind of Finsler gravity is suggested. Physical applications of Finsler decomposed metric are considered. \end{abstract}


\maketitle

\section{Introduction}

Finsler geometry is well known mathematical framework which is a nearest generalization of Riemannian metric geometry~\cite{Rund:1959,Asanov:1985,Matsumoto:1986}. There are various approaches to Finsler formulation of a gravity theory. In Ref. \cite{Chang:2008yv} it has been found that some Finsler structure makes the modified Newton's gravity equivalent to the modified Newtonian dynamics. In Ref. \cite{Huang:2007en} the theory of gravitation in the spacetime with Finsler structure is constructed. In Ref.  \cite{Pfeifer:2011xi} gravitational dynamics for Finsler spacetimes in terms of an action integral on the unit tangent bundle is constructed. In Ref. \cite{Gibbons:2007iu} it was shown that a deformation of very special relativity leads in a natural way to Finsler geometry.

Earlier, Finslerian post-Riemannian corrections to the equations of geodesics have been analyzed in Ref.~\cite{Aringazin-Asanov:1985}.  In Ref.~\cite{Aringazin-Asanov:1987} a class of static even-power Finsler metric functions have been introduced, for which bounds on characteristic Finslerian parameters coming from classical gravitational tests have been obtained. General set up for Finslerian theory of gravitation and Finslerian parameterized post-Newtonian formalism confronted to observable gravitational effects in the Solar system, as well as Finslerian approach to gauge field theory have been presented in Ref.~\cite{Aringazin-Asanov:1988}; Maxwell equations in Finslerian space-time have been obtained from variational principle, and it was shown that in the eikonal approximation these reduce to equations of null geodesics in Finsler space-time. In Ref.~\cite{Aringazin:1989} generalized non-regular Finslerian metrics has been introduced for which all orbital gravitational effects coincide with that implied by the Riemannian general relativity, while the calculated gyroscope precession effect makes a difference.

In Ref. \cite{Lammerzahl:2012kw} a class of spherically symmetric and static Finsler spacetimes which are small perturbations of the Schwarzschild spacetime have been considered. The authors derived equations of motion for freely falling particles and equations for light rays, and discuss the bounds put on the perturbation functions by observations in the Solar system.  In Ref. \cite{Girelli:2006fw} it was shown that the  notion  of Finsler  metric is  the geometric structure encoding the notion of ``rainbow metric''. In Ref.  \cite{Vacaru:2010fi} some generalizations and alternatives to Einstein gravity including modifications with broken local Lorentz invariance were considered. It was also shown how such theories (and general relativity) can be equivalently reformulated in Finsler like variables.

In this paper, we present some approach to Finsler gravity equations that is different from  the above mentioned. With general even-power decomposition anzatz for Finsler metrics, i.e. classes of Finsler metrics which allow decomposition to products of even numbers of Riemannian metrics, being introduced in Ref.~\cite{Aringazin-Asanov:1987}, we consider some special class of Finsler metrics: the metric which can be decomposed as the product of two general Riemannian metrics. In this case one can write Lagrangian with two metrics. Below, we give definitions (Secs. \ref{finsler}, \ref{special}) and then elaborate various procedures to implement this framework (Secs. \ref{pure}, \ref{bimetric}, \ref{general}).

\section{Finsler geometry}
\label{finsler}

Let $M$ is a 4D manifold. The tangent space of $M$ is denoted by $T M$. Suppose $(x^\alpha)$ are local coordinates around $x \in M$. Then we denote the standard basis vectors for $T_x M$ by $y_\alpha = \frac{\partial}{\partial x^\alpha}$, and the standard basis vectors for the cotangent space $T^*_x M$ as $dx^\alpha$.

\textbf{Definition of Finsler metric space}. \textit{Finsler metric space is a manifold $M$ equipped with a function $F : T M > [0, \infty)$ (called Finsler metric function) such that
\begin{enumerate}
  \item $F$ is smooth on $T M \backslash \{0\}$ ;
  \item $F|_{T_xM} : T_x M \rightarrow [0, \infty)$ is a Minkowski norm for all $x \in M$ ;
  \item $F$ is homogeneous of degree one with respect to $y_\alpha$, i.e. $F(x,\lambda y) = \lambda F(x,y)$, $\lambda \not= 0$;
  \item the Hessian of its square, in the fibre coordinates $\dfrac{\partial^2 F}{\partial y^\alpha \partial y^\beta}$, has constant rank and is non-degenerate for all $x,y \in TM$.
\end{enumerate}
}
Finsler metric is defines as
\begin{equation}
\label{0-15}
  ds = F(x,dx).
\end{equation}
\textbf{Definition of Finsler metric tensor.} \textit{Let $F$ be a Finsler metric function on $T M$ and consider $T M$ in the standard induced coordinates. A Finsler metric tensor is the symmetric (0,2) $d-$tensor $g(x, y)$ with components
\begin{equation}
\label{0-10}
  g_{\alpha \beta}(x,y) = \frac{1}{2} \dfrac{\partial^2 F}{\partial y^\alpha \partial y^\beta}.
\end{equation}
}
In the local coordinates, Finsler metric can be written as
\begin{equation}
\label{0-20}
  ds^n = G_{\alpha_1 , \alpha_2 , \cdots , \alpha_n} dx^{\alpha_1} dx^{\alpha_2} \cdots dx^{\alpha_n} = g_{\mu \nu} dx^\mu dx^\nu,
\end{equation}
where $g_{\mu \nu} = G_{\mu \nu; \alpha_1 , \alpha_2 , \cdots , \alpha_{n-2}} dx^{\alpha_1} dx^{\alpha_2} \cdots dx^{\alpha_{n-2}}$.

\section{Special class of Finsler metrics}
\label{special}

Let us consider the special class of Finsler metrics,
\begin{equation}
	ds^4  =
	G_{\alpha \beta\gamma \delta}
	dx^\alpha dx^\beta dx^\gamma dx^\delta = ds_1^2 ds_2^2,
\label{2-10}
\end{equation}
where
\begin{eqnarray}
	ds^2_1 &=& g_{1; \alpha \beta} dx^\alpha dx^\beta,
\label{2-20}\\
	ds^2_2 &=& g_{2; \alpha \beta} dx^\alpha dx^\beta,
\label{2-30}
\end{eqnarray}
i.e.
\begin{equation}
	G_{\alpha \beta\gamma \delta} =
	g_{1; \{\alpha \beta} g_{2; \gamma \delta\}},
\label{2-40}
\end{equation}
where $\{ \}$ means full symmetrization over the indices $\alpha, \beta, \gamma, \delta$.

It is well known that there exist some problems for the definition of dynamical equations for the Finsler metric (analogous to Einstein equations for the standard metrics \eqref{2-20}).

In the present paper, we show that in the special choice of Finsler metric \eqref{2-10} (we called it decomposed Finsler metric) it is possible to introduce well defined gravitational Finsler equations. The main idea is to use two Riemannian metrics $ds^2_{1,2}$ in the definition of Lagrangian and corresponding field equations. After that we can use well known bimetric formalism for constructing corresponding Lagrangian and field equations.

\section{Pure geometrical gravity for decomposed Finsler metric}
\label{pure}

One can construct many different pure geometrical Finsler gravities. Below, we would like to present local and nonlocal models.

\subsection{Local pure geometrical decomposed Finsler gravity}

In this approach we construct pure geometrical Lagrangian $\mathcal L$ as follows
\begin{equation}\label{3e-10}
  \mathcal L = - \dfrac{1}{2 \varkappa} \left(
  k_1 R_1 \sqrt{-g_2} + k_2 R_2 \sqrt{-g_1}
  \right) + \mathcal L_m
\end{equation}
where $\varkappa$ is Einstein gravitational constant; $k_{1,2}$ are constants; $R_{1,2}$ are scalar curvatures for $g_{1,2}$ metrics and $\mathcal L_m$ is Lagrangian density for a matter. Using Palatini formalism we will obtain following equations. Varying with respect to the metrics components
$g_{1,2; \mu \nu}$ gives us
\begin{eqnarray}
\label{3e-20}
  k_1 R_{1; \mu \nu} - k_2 R_2 \frac{1}{2} g_{1; \mu \nu} &=&
  \varkappa T_{1; \mu \nu} ,
\\
\label{3e-30}
  k_2 R_{2; \mu \nu} - k_1 R_1 \frac{1}{2} g_{2; \mu \nu} &=&
  \varkappa T_{2; \mu \nu}
\end{eqnarray}
where $T_{1,2; \mu \nu}$ are energy momentum tensors obtained varying with respect to metrics $g_{1,2}$.

Varying with respect to the affine connection $\Gamma^\alpha_{\beta \gamma}$ and using standard mathematical manipulations due to Palatini approach gives us the equation determining the form of the connection by the metric tensor,
\begin{equation}
\label{3e-40}
  \Gamma_{i; \lambda \sigma}^\mu = \dfrac{1}{2} g_i^{\gamma \mu}
  \left(
    \Delta_{i; \gamma \lambda \sigma} + \Delta_{i; \gamma \sigma \lambda} -
    \Delta_{i; \sigma \lambda \gamma}
  \right), \quad i=1,2,
\end{equation}
where $\Delta_{i; \lambda \gamma \sigma}$ is
\begin{equation}\label{3e-50}
\begin{split}
  \Delta_{i; \lambda \gamma \sigma} = &
  \dfrac{\partial g_{i; \lambda \gamma}}{\partial x^\sigma} +
  g_{i; \lambda \gamma} \left[
    \frac{1}{\sqrt{-g_j}} \left(\sqrt{-g_j}\right)_{, \sigma} -
    \frac{1}{\sqrt{-g_i}} \left(\sqrt{-g_i}\right)_{, \sigma}
  \right] =
\\
  &
  \dfrac{\partial g_{i; \lambda \gamma}}{\partial x^\sigma} +
  g_{i; \lambda \gamma} \left(
    \ln \sqrt{\dfrac{g_j}{g_i}}
  \right)_{, \sigma}
  ,  \quad i+j=2.
\end{split}
\end{equation}
We see that such interpretation of Finsler geometry gives rise to the affine connection that is not compatible with the metric.

\subsection{Nonlocal pure geometrical Finsler gravity}
\label{nonlocal}

Nonlocal version of a pure geometrical Finsler gravity can be defined in the following way:
\begin{equation}\label{4b-10}
  S = \int R_1(x_1) G(x_1, x_2) R_2(x_2) \sqrt{-g_1} \sqrt{-g_2} \;
  dx_1 dx_2
\end{equation}
where $S$ is the action; $R_{1,2}$ are the same as for equation \eqref{3e-10} and $G(x_1, x_2)$ is some nonlocal function.

\section{Bimetric interpretation of decomposed Finsler metric}
\label{bimetric}

In this section, we use the bimetric formalism to derive dynamical equations for the metric geometry.

\subsection{$ds_1^2$ as the metric for gravity and $ds_2^2$ as the metric for matter}

In this subsection we use the following interpretation of the decomposed Finsler metric. The first metric $ds_1^2$ is used as the metric for gravity and the second metric $ds_2^2$ as the metric for matter (for the corresponding bi-metric formalism see Ref. \cite{Drummond:2001rj}).

Let us introduce two vierbein bundles into the space-time manifold. Each bundle supports its own metric. One is associated with underlying gravity and the other with matter.

The vierbein bundle appropriate to gravity is $e_{\mu a}$, with the associated metric
\begin{equation}
	ds_1^2 = e_{\mu a} e_\nu^{\phantom{\nu} a}
\label{3a-10}
\end{equation}
where  raising and lowering of vierbein indices $a$ are carried out with the standard Minkowski metric $\eta_{ab} = \left\{ +1, -1, -1, -1 \right\}$.
The vierbein associated with matter is $\bar e_{\mu a}$ and the raising and lowering of $\bar a$-indices are made by means of the Minkowski metric
$\bar \eta_{ab} = \eta_{ab}$. The associated metric is
\begin{equation}
	ds_2^2 = \bar e_{\mu \bar a} \bar e_\nu^{\phantom{\nu} \bar a} .
\label{3a-20}
\end{equation}
The two vierbein bundles are related to each other by some local Lorentz transformation $M^a_{\phantom{a} \bar a} \in SL(4,R)$ and the factor $e^\phi$ is the local scaling.

The dynamics is derived from the action which we represent in three parts, the gravitational term $S_g$ corresponding to the metric $ds_1^2$, the matter term
$S_m$ corresponding to the matter metric $ds_2^2$, and the linking action $S_L$, which depends on the variables that determine the relationship between the two vierbein bundles. The full action $I$ is the sum of the three terms,
\begin{equation}
	S = S_g + S_M + S_L .
\label{3a-30}
\end{equation}
The gravitational action has standard form
\begin{equation}
	S_g = - \frac{1}{4 \pi G} \int R \sqrt{-g} \ d^4 x
\label{3a-40}
\end{equation}
where $G$ is the coupling constant having the same dimensionality as Newton's constant $G_N$. The linking action is
\begin{equation}
	S_L = \frac{1}{4 \pi F} \int g^{\mu \nu} \mathrm{Tr} \left(
		j_\mu j_\nu
	\right)
	\sqrt{-g} \ d^4 x + \frac{1}{4 \pi F'} \int g^{\mu \nu} \left(
		\partial_\mu \phi \partial_\nu \phi
	\right) \sqrt{-g} \ d^4 x
\label{3a-50}
\end{equation}
where $F$ and  $F'$ are new gravitational constants. The matrix valued current $j_\mu$ is given by
\begin{equation}
	j_\mu = \left( D_\mu M \right) M^{-1},
\label{3a-60}
\end{equation}
where $D_\mu$ is the covariant derivative. The matter action is given by
\begin{equation}
	S_M = \int L_M(\psi_A; \bar e_{\mu \bar a}) \sqrt{-g} \ d^4 x,
\label{3a-70}
\end{equation}
where $\psi_A$ is matter; $A$ is collective index and
$\bar e_{\mu \bar a}$ is the vierbein desribing matter.

\subsection{Finsler MOND gravity}

In this subsection, we apply two parts \eqref{2-20} and \eqref{2-30} of the Finsler metric \eqref{2-10} to address the problem of describing dark matter. We follow the approach developed in Ref. \cite{Milgrom:2009gv}. Due to this approach, one can form nontrivial tensors and scalars using the difference of two corresponding Levi-Civita connections $\Gamma^\alpha_{\beta \gamma}$ and $\hat \Gamma^\alpha_{\beta \gamma}$,
\begin{equation}
	C^\alpha_{\beta \gamma} = \Gamma^\alpha_{\beta \gamma} -
	\hat \Gamma^\alpha_{\beta \gamma}.
\label{3b-10}
\end{equation}
Introducing two covariant derivatives denoted by $(;)$ with respect to the connection $\Gamma^\alpha_{\beta \gamma}$ and $(:)$ with respect to the connection
$\hat \Gamma^\alpha_{\beta \gamma}$, we obtain the following relationships between metrics and connections:
\begin{eqnarray}
	g_{\mu \nu : \lambda} &=& g_{\alpha \nu} C^\alpha_{\mu \lambda} +
	g_{\alpha \mu} C^\alpha_{\nu \lambda} , \quad
	\hat g_{\mu \nu ; \lambda} = - \hat g_{\alpha \nu} C^\alpha_{\mu \lambda} -
	\hat g_{\alpha \mu} C^\alpha_{\nu \lambda} ,
\label{3b-20}\\
	C^\lambda_{\alpha \beta} &=& \frac{1}{2} g^{\lambda \rho} \left(
		g_{\alpha \rho : \beta} + g_{\beta \rho : \alpha} -
		g_{\alpha \beta : \rho}
		\right) = - \frac{1}{2} \hat g^{\lambda \rho} \left(
		\hat g_{\alpha \rho ; \beta} + \hat g_{\beta \rho ; \alpha} -
		\hat g_{\alpha \beta ; \rho}
	\right).
\label{3b-30}
\end{eqnarray}
We introduce the tensor
\begin{equation}
	\Upsilon_{\mu \nu} = C^\gamma_{\mu \lambda} C^\lambda_{\nu \gamma} -
	C^\gamma_{\mu \nu} C^\lambda_{\lambda \gamma}.
\label{3b-40}
\end{equation}
and the following scalars and tensors:
\begin{eqnarray}
	R_{\mu \nu} &=& \Gamma^\alpha_{\mu \alpha, \nu} -
	\Gamma^\alpha_{\mu \nu, \alpha} +
	\Gamma^\gamma_{\mu \lambda} \Gamma^\lambda_{\nu \gamma} -
	\Gamma^\gamma_{\mu \nu} \Gamma^\lambda_{\lambda \gamma},
\label{3b-50}\\
	\hat R_{\mu \nu} &=& \hat \Gamma^\alpha_{\mu \alpha, \nu} -
	\hat \Gamma^\alpha_{\mu \nu, \alpha} +
	\hat \Gamma^\gamma_{\mu \lambda} \hat \Gamma^\lambda_{\nu \gamma} -
	\hat \Gamma^\gamma_{\mu \nu} \hat \Gamma^\lambda_{\lambda \gamma},
\label{3b-60}\\
	R &=& g^{\mu \nu} R_{\mu \nu},
\label{3b-70}\\
	\hat R &=& \hat g^{\mu \nu} \hat R_{\mu \nu},
\label{3b-80}\\
	\Upsilon &=& g^{\mu \nu} \Upsilon_{\mu \nu},
\label{3b-90}\\
	\hat R_m &=& g^{\mu \nu} \hat R_{\mu \nu},
\label{3b-100}\\
	R_m &=& \hat g^{\mu \nu} R_{\mu \nu}.
\label{3b-110}
\end{eqnarray}
Now one can construct gravitational Lagrangian densities using the scalars
$R$, $\hat R$, $\Upsilon$, $\hat R_m$, and  $R_m$, as well as scalars constructed by
contracting $C^\lambda_{\mu \nu}$ with the two metrics and their reciprocals. For completeness, we note that the entities $\hat g/g, \bar \omega = g^{\mu \nu} \hat g_{\mu \nu}$, etc.  can be used for this purpose as well.

In Ref. \cite{Milgrom:2009gv} it was pointed out that there exists such a class of relativistic theories that have MOND-like theories in a non-relativistic limit, and which produce enhanced, MOND-like gravitational lensing.

\subsection{Finsler geometry and variable Speed of Light Cosmology}

One can relate two metrics \eqref{2-20} and \eqref{2-30} of spacetime in the following way \cite{Moffat:2004qs}:
\begin{equation}
	\hat g_{\mu \nu} = g_{\mu \nu} + B \partial_\mu \phi \partial_\nu \phi .
\label{3c-10}
\end{equation}
Here, metric $\hat g_{\mu \nu}$ is called ``matter'' metric and
$g_{\mu \nu}$ is gravitational metric. Corresponding Lagrangian consists of a scalar field coupled to matter through the matter metric
$\hat g_{\mu \nu}$  with the action
\begin{equation}
  S = S_{\text{grav}} + S_{\phi} + \hat{S}_m,
\label{3c-20}
\end{equation}
where
\begin{equation}
  S_{\text{grav}} = - \frac{1}{\kappa} \int
  \left( R[g] + 2 \Lambda \right) \sqrt{-g} \ d\Omega.
\label{3c-30}
\end{equation}
Here, $\Lambda$ is the cosmological constant and $c$ denotes the currently measured speed of light.

In the bimetric gravity one can choose either ${\hat g}_{\mu\nu}$ or $g_{\mu\nu}$ to be comoving metric frame in the FRW universe (here we follow Ref. \cite{Moffat:2004qs}). There are two characteristic speeds: the speed of light $c_\gamma$ and the speed of gravitational waves $c_g$. If we choose ${\hat g}_{\mu\nu}$ as the comoving metric, then
\begin{equation}
  d{\hat s}^2 \equiv {\hat g}_{\mu\nu}dx^\mu dx^\nu = c^2 dt^2 - R^2(t)
  \left[ \frac{dr^2}{1-kr^2} +
    r^2(d\theta^2 + \sin^2\theta d\phi^2)
  \right],
\label{3c-40}
\end{equation}
and
\begin{equation}
  ds^2 \equiv g_{\mu\nu}dx^\mu dx^\nu = c^2
  \left( 1-\frac{B}{c^2}\dot\phi^2
  \right) dt^2 - R^2(t)
  \left[ \frac{dr^2}{1-kr^2} +
    r^2(d\theta^2 + \sin^2\theta d\phi^2)
  \right].
\label{3c-50}
\end{equation}
From \eqref{3c-50} we see that the speed of gravitational waves is
given by $c_g(t)=c\left( 1-\frac{B}{c^2}\dot\phi^2 \right)^{1/2}$, while the speed of light is constant. Alternatively, if we choose $g_{\mu\nu}$ as the comoving metric, we obtain
\begin{equation}
  ds^2 \equiv g_{\mu\nu}dx^\mu dx^\nu = c_g^2 dt^2 - R^2(t)
  \left[
  \frac{dr^2}{1 - kr^2} + r^2
  \left( d\theta^2 + \sin^2\theta d\phi^2 \right)
  \right]
\label{3c-60}
\end{equation}
and
\begin{equation}
  d{\hat s}^2 \equiv {\hat g}_{\mu\nu}dx^\mu dx^\nu  = c^2 \left(1+\frac{B}{c^2}\dot\phi^2 \right)dt^2 - R^2(t)
  \left[
  \frac{dr^2}{1-kr^2} + r^2
  \left( d\theta^2 + \sin^2\theta d\phi^2 \right)
  \right].
\label{3c-70}
\end{equation}
Now the speed of light is given by
$c_\gamma(t) = c \left( 1+\frac{B}{c^2}\dot\phi^2 \right)^{1/2}$, while the speed of gravitational waves $c_g=c$ is constant. One can show that by regauging clocks, we cannot make both $c_\gamma$ and $c_g$ simultaneously constant. This makes the time dependence of either $c_\gamma$ or $c_g$ a non-trivial feature of the bimetric gravity.

In Ref. \cite{Moffat:2004qs} it was shown that the variable speed of light  mechanism in the bimetric gravity model can solve the flatness and horizon problems and it is possible that one can avoid the initial fine-tuning problems of generic inflationary models.

\subsection{Finsler geometry and massive gravity}

Dynamics with two metric can be defined by the action \cite{Visser:1997hd}
\begin{equation}
	S = \int d^4 x \left\{
		\frac{1}{16 \pi} \left[
			\sqrt{-g} R + \sqrt{-b} L_{mass}(g, g_0) +
			\sqrt{-g} L_{matter} (g, X)
		\right]
	\right\}
\label{3d-10}
\end{equation}
where $g_{0,\mu \nu}$ represents the background metric tensor ($ds_1^2$ metric in our formalism), $g_{\mu \nu}$ is the dynamical metric tensor ($ds_2^2$ metric in our formalism), and $X$ stands for any nongravitational field.

The theory with such Lagrangian is used for the explanation of an acceleration of the Universe in modern epoch \cite{Visser:1997hd}.

The modified Einstein equations thus are
\begin{equation}
  R^{\mu\nu} - \frac{1}{2} g^{\mu \nu} R =
  8\pi G \left( T^{\mu\nu}_{mass} + T^{\mu\nu} \right),
\label{3d-20}
\end{equation}
where $T^{\mu\nu}_{mass}$ is an extra contribution to the stress-energy tensor,
\begin{equation}
  T_{mass}^{\mu\nu} =
  - \dfrac{m_g^2 c^2}{8\pi G \hbar^2}
  \left\{
    (g_0^{-1})^{\mu\sigma} \;
    \left[ (g- g_0)_{\sigma\rho} - {1\over2} (g_0)_{\sigma\rho} \;
    (g_0^{-1})^{\alpha\beta} (g- g_0)_{\alpha\beta} \right]
    (g_0^{-1})^{\rho\nu}
  \right\}.
\label{3d-30}
\end{equation}
This kind of gravity has the following features:
\begin{itemize}
  \item black holes (of the usual type) do not exist in this theory;
  \item the expansion of the universe can be completely divorced from the cosmological distribution of matter.
\end{itemize}

\section{Non bimetric case}
\label{general}

In this section, we discuss more general case. Let us consider the Finsler metric \eqref{2-10} where $G_{\alpha,\beta,\gamma,\delta}$ does not allow the decomposition into two metrics as in \eqref{2-40}. We will try the case when it can be decomposed using four metrics in the following way:
\begin{equation}
	ds^4 =
	G_{\alpha \beta\gamma \delta} = \sum\limits_{i,j=1}^4 ds_i^2 ds_j^2
\label{5-10}
\end{equation}
where for any $i=1,2,3,4$ metric $ds_i^2$ is represented as
\begin{equation}
	ds_i^2 = g_{i; \mu \nu} dx^\mu dx^\nu .
\label{5-20}
\end{equation}
First of all it is necessary to solve the equations
\begin{equation}
	G_{\alpha \beta \gamma \delta} = \sum\limits_{i,j=1}^4
	g_{i; \{ \alpha \beta} g_{j; \gamma \delta \}}
\label{5-30}
\end{equation}
where $\{ \alpha \beta \gamma \delta \}$ is the symmetrization over $\alpha,\beta,\gamma,\delta$. In order to consider the solvability of this equation, we calculate the number of independent elements of $G_{\alpha \beta \gamma \delta}$ and four matrices $g_{i; \mu \nu}$. Let us remind that $G_{\alpha \beta \gamma \delta}$ is symmetric over all indices.

The tensor $G_{\alpha \beta \gamma \delta}$ has
\begin{itemize}
	\item all indices $\alpha \beta \gamma \delta$ are different: 1;
	\item two pairs with equal indices, $G_{\alpha \alpha \beta \beta}$: 10;
	\item three equal indices: 16;
	\item all indices are equal: 4.
\end{itemize}
Thus we have 31 independent components of $G_{\alpha \beta \gamma \delta}$. Four metrics $g_{i; \mu \nu}$ have in total 40 independent components. Consequently, 9 degrees of freedom can be chosen in an arbitrary way.

For each of the metric $g_{i; \mu \nu}$ one can introduce appropriate covariant derivative $\stackrel{i}{\nabla}_\mu, i=1,2,3,4$. Then one can introduce the following associated geometric entities:
\begin{eqnarray}
	(R_i)_{\mu \nu} &=& (\Gamma_i)^\alpha_{\mu \alpha, \nu} -
	(\Gamma_i)^\alpha_{\mu \nu, \alpha} +
	(\Gamma_i)^\gamma_{\mu \lambda} (\Gamma_i)^\lambda_{\nu \gamma} -
	(\Gamma_i)^\gamma_{\mu \nu} (\Gamma_i)^\lambda_{\lambda \gamma},
\label{5-40}\\
	R_i &=& g_i^{\mu \nu} (R_i)_{\mu \nu},
\label{5-50}\\
	(C_{ij})^\lambda_{\alpha \beta} &=& \frac{1}{2} g_i^{\lambda \rho} \left(
		\stackrel{j}{\nabla_\beta} g_{\alpha \rho} +
		\stackrel{j}{\nabla_\alpha} g_{\beta \rho} -
		\stackrel{j}{\nabla_\rho} g_{\alpha \beta}
		\right),
\label{5-60}\\
	(\Upsilon_{ij})_{\mu \nu} &=& (C_{ij})^\gamma_{\mu \lambda}
	(C_{ij})^\lambda_{\nu \gamma} -
	(C_{ij})^\gamma_{\mu \nu} (C_{ij})^\lambda_{\lambda \gamma},
\label{5-80}\\
	\Upsilon_{ijk} &=& g_k^{\mu \nu} (\Upsilon_{ij})_{\mu \nu},
\label{5-90}\\
	(R_{ij}) &=& g_i^{\mu \nu} (R_j)_{\mu \nu},
\label{5-100}\\
	\omega_{ij} &=& g_i^{\mu \nu} g_{j; \mu \nu}
\label{5-110}
\end{eqnarray}
where $(R_i)_{\mu \nu}$ is the Ricci tensor for the metric $g_{i; \mu \nu}$. Using the scalars $\Upsilon_{ijk}$, $R_{ij}$, etc. one can write Lagrangian. Then, we define the action with terms where integrating is made over all $g_i = \det g_{i; \mu \nu}$. Below, we present examples of the models of pure geometrical Finsler gravities.

\subsection{Local pure geometrical decomposed Finsler gravities}

First, we can define Lagrangian in the following local form:
\begin{equation}\label{5b-10}
  \mathcal L = \sum \limits_{i,j} k_{i} R_{i} \sqrt{-g_j} +
  \sum \limits_{i,j,k} k_{ij} R_{ij} \sqrt{-g_k} +
  (\text{various type of scalars from \eqref{5-80}-\eqref{5-110}}),
\end{equation}
where $k_i$ and $k_{ij}$ are constants.

\subsection{Nonlocal pure geometrical decomposed Finsler gravities}

Second, we can also define action in the following nonlocal form:
\begin{equation}\label{5c-10}
\begin{split}
  S = & \int \left(
    R_i(x_i) G(x_i, x_j) R_j(x_j) \sqrt{-g_i} \sqrt{-g_j} \;
    dx_i dx_j +
    \right.
\\ 
  & \left.
    R_i(x_i) G(x_i, x_j, x_k) R_j(x_j) R_j(x_k)
    \sqrt{-g_i} \sqrt{-g_j} \sqrt{-g_k} \;
    dx_i dx_j dx_k + \cdots
  \right),
\end{split}
\end{equation}
where $G(x_i, x_j), G(x_i, x_j, x_k), \ldots$ are nonlocal functions.

\section{Conclusions}

We have shown that for some special choice of Finsler metric one can obtain a new particular formulation of Finsler gravity. The main point of this formulation is that we specify the class of Finsler metrics defined by the condition that they allow decomposition to the product of two general Riemannian metrics. In this case the Finsler metric is thought of as equivalent to two Riemannian metrics (Sec. \ref{special}).

In section \ref{pure} we present new formulation of Finsler gravity for our class of Finsler metric. In section \ref{bimetric} we present and discuss various physical applications of this special form of Finsler gravity. As one of the results we obtained that in this case the affine connection becomes non-compatible with the metric. The difference is given by the gradient $(\ln \sqrt{g_1/g_2})_{, \mu}$.

Also, in section \ref{general} we constructed local and nonlocal Finsler gravities arised from the more general case, namely, when Finsler metrics allows decomposition to four Riemannian metrics.

\section*{Acknowledgments}

This work was partially supported  by a grant in fundamental research in natural sciences by Science Committee of the Ministry of Education and Science of Kazakhstan and by a grant of VolkswagenStiftung.

\end{document}